# Defect driven flexo-chemical coupling in thin ferroelectric films

*Eugene A. Eliseev,[1] Ivan. S. Vorotiahin[2,3], Yevhen M. Fomichov [1,4], Maya D. Glinchuk[1], Sergei V. Kalinin[5], Yuri A. Genenko[3*], and Anna N. Morozovska,[2†]*

[1]Institute for Problems of Materials Science, National Academy of Sciences of Ukraine,

3, Krjijanovskogo, 03142 Kyiv, Ukraine

[2]Institute of Physics, National Academy of Sciences of Ukraine,

46, Prospekt Nauky, 03028 Kyiv, Ukraine

[3] Institut für Materialwissenschaft, Technische Universität Darmstadt, Jovanka-Bontschits-Str. 2, 64287 Darmstadt, Germany

[4] Faculty of Mathematics and Physics, Charles University, V Holesovickach 2, 18000 Prague 8, Czech Republic

[5] The Center for Nanophase Materials Sciences, Oak Ridge National Laboratory,

Oak Ridge, TN 37831

**Abstract**

Using Landau-Ginzburg-Devonshire theory, we considered the impact of the flexoelectro-chemical coupling on the size effects inpolar properties and phase transitions of thin ferroelectric films with a layer of elastic defects. We investigated a typical case, when defects fill a thin layer below the top film surface with a constant concentration creating an additional gradient of elastic fields. The defective surface of the film is not covered with an electrode, but instead with an ultra-thin layer of ambient screening charges, characterized by a surface screening length. This geometry is typical for the scanning probe piezoelectric force microscopy.

Obtained results revealed an unexpectedly strong effect of the joint action of Vegard stresses and flexoelectric effect (shortly flexo-chemical coupling) on the ferroelectric transition temperature, distribution of the spontaneous polarization and elastic fields, domain wall structure and period in thin $PbTiO_3$ films containing a layer of elastic defects. A nontrivial result is the ferroelectricity persisting at film thicknesses below 4 nm, temperatures lower than 350 K and relatively high surface screening length (~0.1 nm). The origin of this phenomenon is the re-building of the domain structure in the film (namely the cross-over from c-domain stripes to a-type closure domains) when its thickness decreases below 4 nm, conditioned by the flexoelectric coupling and facilitated by negative Vegard effect. For positive Vegard effect, thicker films exhibit the appearance of pronounced maxima on the thickness dependence of the transition temperature, whose position and height can be controlled by the defect type and concentration. The revealed features may have important implications for miniaturization of ferroelectric-based devices.

---

[*] Corresponding author 2. E-mail: genenko@mm.tu-darmstadt.de
[†] Corresponding author 1. E-mail: anna.n.morozovska@gmail.com

# I. INTRODUCTION

Deep physical understanding and a possible control of the thin ferroelectric films polar properties are important for both fundamental research and the most promising applications in memory elements as well as many other devices [1]. With decreasing the film thickness its ferroelectric properties usually decline until their complete disappearance at thicknesses smaller than the critical one [2]. Feasible ways to avoid the size-induced phase transition in thin epitaxial films are, for example, selecting of an appropriate substrate [3] or modification of their chemical composition [4]. Particularly it was shown that the retaining of ferroelectricity down to ultrathin films (3 – 5 lattice constants thick) is provided by the "self-polarizing" role of elastic strains arising in the film due mismatching lattice constants of the film and the substrate [5]. The presence of point elastic defects (such as uncharged impurities and vacancies, elastic dipoles, dilatation centers [6]) can strongly impact the electric polarization of films via electrostriction [7], "chemical" strains or Vegard stresses (see e.g. [8, 9, 10, 11]), and flexoelectric effect (see e.g. [12, 13, 14, 15]). Due to the gradient nature, the elastic defect influence is much more complex and less studied [16] than the effect of homogeneous elastic strains arising in a film due to the film and substrate lattice mismatch. A joint action of Vegard stresses and flexoelectric effect, named **flexo-chemical effect** [17], can explain some unusual phenomena caused by size effects, such as e.g. reentrant ferroelectric phase with enhanced polarization at room temperature observed in BaTiO$_3$ nanoparticles with sizes less than 20 nm [18].

However, as one can see from **Table I,** the influence of combined flexo-chemical and size effects on domain structures, polar, elastic and electrophysical properties of thin ferroelectric films was not considered so far. The main objective of our work is to advance a theory describing the impact of **defect driven** flexo-chemical coupling on the film properties and size effects, and analyze the outcomes towards optimization of the properties for advanced applications.

Table I. Flexoelectric, Vegard and size effects, domain structure formation considered ("**Yes**" or "**No**") in ferroelectric thin films

| References examples | Flexoelectric, Vegard and size effects, domain structure formation considered (Yes or No) in ferroelectric thin films | | | | | | |
| --- | --- | --- | --- | --- | --- | --- | --- |
| | Flexo-electric effect | gradient, of compo-sition | Composi-tional polariza-tion | Vegard stresses | Flexo-chemical effect | Domain formation | Size effects |
| Tagantsev et al. [19] and refs. therein | No | N/A | N/A | No | No | No | Yes |
| Tilley [2] | No | No | No | No | No | Yes | Yes |
| Catalan et al [20] | Yes | Yes | No | No | No | No | Yes |
| Marvan et al.[21] | No | Yes | Yes | No | No | No | No |
| Bratkovsky, and | No | Yes | Yes | No | No | No | Yes |
| Levanyuk [22] | | | | | | | |
| Ban et al [23], Zhong et al [24] | No | Yes | No | No | No | No | No |
| Karthik et al. [25] | Yes | Yes | No | No | No | No | No |
| Morozovska et al [26] | No | Yes | No | Yes | No | Yes | No |
| Morozovska et al [16] | Yes | Yes | No | Yes | No | No | Yes |
| Tagantsev and Yudin [27] and refs therein | Yes | N/A | No | No | No | Yes | No |
| Morozovska et al [28], | Yes | Yes | No | Yes | No | No | Yes |
| Vorotiahin et al. [29] | Yes | No | No | No | No | No | Yes |
| Vorotiahin et al. [30] | Yes | No | No | No | No | Yes | Yes |
| **This work** | Yes | Yes | Yes | Yes | Yes | Yes | Yes |

## II. STATEMENT OF THE PROBLEM

The Landau-Ginzburg-Devonshire (LGD) expansion of bulk ($G_V$) and surface ($G_S$) parts of the Gibbs free energy of a ferroelectric film in powers of the polarization vector and stress tensor components $P_i$ and $\sigma_{ij}$ and the energy of the electric field outside the film ($G_{ext}$) have the form:

$$G = G_V + G_S + G_{ext}, \quad (1a)$$

$$G_V = \int_{V_{FE}} d^3r \left( \begin{array}{c} \frac{a_{ik}}{2} P_i P_k + \frac{a_{ijkl}}{4} P_i P_j P_k P_l + \frac{a_{ijklmn}}{6} P_i P_j P_k P_l P_m P_n + \frac{g_{ijkl}}{2}\left(\frac{\partial P_i}{\partial x_j}\frac{\partial P_k}{\partial x_l}\right) - P_i E_i \\ -\frac{\varepsilon_0 \varepsilon_b}{2} E_i E_i - \frac{s_{ijkl}}{2}\sigma_{ij}\sigma_{kl} - Q_{ijkl}\sigma_{ij}P_k P_l - F_{ijkl}\left(\sigma_{ij}\frac{\partial P_l}{\partial x_k} - P_l\frac{\partial \sigma_{ij}}{\partial x_k}\right) - W_{ij}\sigma_{ij}\delta N \end{array}\right), \quad (1b)$$

$$G_S = \int_S d^2r \left( \frac{a_{ij}^S}{2} P_i P_j - \frac{\varepsilon_0}{2\lambda}\varphi^2 \right), \qquad G_{ext} = -\int_{\vec{r}\notin V_{FE}} d^3r \frac{\varepsilon_0 \varepsilon_e}{2} E_i E_i. \quad (1c)$$

The tensor $a_{ij}$ is positively defined for linear dielectrics, and explicitly depends on temperature $T$ for ferroelectrics and paraelectrics. Below we use an isotropic approximation for the tensor coefficients $a_{ij}^S = \alpha_{S0}\delta_{ij}$ and $a_{ij} = \alpha_T(T - T_c)\delta_{ij}$, where $\delta_{ij}$ is the Kroneker delta symbol, $T$ is absolute temperature, $T_c$ is the Curie temperature. All other tensors included in the free energy (1) are supposed to be temperature independent. Tensor $a_{ijklmn}$ should be positively defined for the thermodynamic stability in paraelectrics and ferroelectrics. Tensor $g_{ijkl}$ determines the magnitude of the gradient energy, and is also regarded positively defined. $\varepsilon_0$ is the vacuum permittivity, $\varepsilon_b$ is a relative background dielectric permittivity [31]. Coefficients $Q_{ijkl}$ are the components of electrostriction tensor; $s_{ijkl}$ are the components of elastic compliance tensor, $F_{ijkl}$ is the flexoelectric strain coupling tensor. For most of the cases one can neglect the polarization relaxation and omit high order elastic strain gradient terms if the flexoelectric coefficients are below the critical values $F_{ijkl}^{cr}$ [32, 33]. $W_{ij}$ is the elastic dipole (or Vegard strain) tensor, that is regarded diagonal hereinafter, i.e. $W_{ij} = W\delta_{ij}$. The quantity $\delta N = N(\vec{r}) - N_e$ is the difference between the concentration of defects $N(r)$ at the point **r** and their equilibrium (average) concentration $N_e$. Here we introduce electric field via electrostatic potential φ as $E_i = -\partial\varphi/\partial x_i$ and an effective surface screening length λ that can be much smaller than lattice constant [34]. Polarization is conjugated to the electric field $E_i$ which can include external and depolarization contributions (if any exists).

Note that we neglected the higher elastic gradient term $\frac{1}{2}v_{ijklmn}(\partial\sigma_{ij}/\partial x_m)(\partial\sigma_{kl}/\partial x_n)$ in the functional (1b), because its magnitude and sign are still disputed [35]. Thus we apply only one half ($F_{ijkl}P_k(\partial\sigma_{ij}/\partial x_l)$) of the full Lifshitz invariant $F_{ijkl}(P_k(\partial\sigma_{ij}/\partial x_l) - \sigma_{ij}(\partial P_k/\partial x_l))/2$. The higher elastic gradient term is necessary for the stability of the thermodynamic potential if the full Lifshitz invariant is included. Application of either the term $F_{ijkl}P_k(\partial\sigma_{ij}/\partial x_l)$ or the term $F_{ijkl}(P_k(\partial\sigma_{ij}/\partial x_l) - \sigma_{ij}(\partial P_k/\partial x_l))/2$ results in the same equations of state. The full form, however, leads to the higher order elastic equations and affects the boundary conditions [36, 37, 38, 39]. The reason of using only the part of the Lifshitz invariant in Eq. (1) is that implementation of the full form causes poor convergence of the numerical code and impairs the quality and reliability of the obtained results. Using the truncated form in Eq. (1) can be justified by the smallness of the flexoelectric coupling strength as compared to the polarization gradient term. Thus following Refs. [40, 41] we assume that the used approximation is valid if $F_{klmn}^2 \ll g_{ijkl}s_{ijmn}$.

Polarization distribution can be found from the Euler-Lagrange equations obtained after variation of the free energy (1)

$$a_{ik}P_k + a_{ijkl}P_jP_kP_l + a_{ijklmn}P_jP_kP_lP_mP_n - g_{ijkl}\frac{\partial^2 P_k}{\partial x_j \partial x_l} - Q_{ijkl}\sigma_{kl}P_j + F_{ijkl}\frac{\partial\sigma_{kl}}{\partial x_j} = E_i, \quad (2a)$$

along with the boundary conditions on the top surface of the film $S$ at $x_3 = h$:

$$\left(g_{kjim}n_k\frac{\partial P_m}{\partial x_j} + a_{ij}^S P_j - F_{jkim}\sigma_{jk}n_m\right)\bigg|_{x_3=h} = 0. \quad (2b)$$

The most evident consequences of the flexocoupling are the inhomogeneous terms in the boundary conditions (2b).

Elastic stress tensor satisfies the mechanical equilibrium equation $\partial\sigma_{ij}/\partial x_j = 0$; elastic strains are $u_{ij} = -\delta G_V/\delta\sigma_{ij}$. The boundary conditions at the mechanically free surface $x_3 = h$ can be obtained from the variation of the free energy (1) with respect to the stresses:

$$\sigma_{ij}n_j\big|_S = 0. \quad (3a)$$

Here $n_j$ are components of the external normal to the film surface. Misfit strain $u_m$ existing at the film-substrate interface ($x_3 = 0$) leads to the boundary conditions for mechanical displacement components, $U_i$ related to elastic strain as $u_{ij} = (\partial U_i/\partial x_j + \partial U_j/\partial x_i)/2$:

$$(U_1 - x_1 u_m)|_{x_3=0} = 0, \quad (U_2 - x_2 u_m)|_{x_3=0} = 0, \quad U_3|_{x_3=0} = 0. \tag{3b}$$

The periodic conditions were imposed at the lateral sides, $U_1|_{x_1=-w/2} - U_1|_{x_1=w/2} = wu_m$, $U_2|_{x_2=-w/2} - U_2|_{x_2=w/2} = wu_m$, while the period $w$ should be defined self-consistently.

The electric field **E** (being the sum of an external $\mathbf{E}^{ext}$ and a depolarization one $\mathbf{E}^d$) is determined self-consistently from the electrostatic problem for the electric potential φ,

$$\varepsilon_0 \varepsilon_b \frac{\partial^2 \varphi}{\partial x_i \partial x_i} = -\frac{\partial P_j}{\partial x_j}, \tag{4}$$

supplemented by the condition of potential continuity at the top surface of the film, $z = h$., using hereinafter notations $x_1 \equiv x, x_2 \equiv y, x_3 \equiv z$. The difference of electric displacement components $D_n^{(i)} - D_n^{(e)}$ is conditioned by the surface screening produced by the ambient free charges at the film surface $S$:

$$\left(\varphi^{(e)} - \varphi^{(i)}\right)\Big|_{x_3=h} = 0, \quad \left(D_n^{(i)} - D_n^{(i)} + \varepsilon_0 \frac{\varphi}{\lambda}\right)\Big|_{x_3=h} = 0 \tag{5}$$

Here **n** is the outer normal to the film surface, electric displacement $\mathbf{D} = \varepsilon_0 \varepsilon_b \mathbf{E} + \mathbf{P}$, the subscript "$i$" means the physical quantity inside the film, "$e$" – outside the film. The conditions of zero potentials were imposed at the bottom electrode ($z = 0$) and a remote top electrode ($z = H + h$, $H \to \infty$), respectively [42] (see **Fig. 1)**.

We suppose that most of defects are located in a thin top layer of thickness $h_0$ beyond which their concentration decreases exponentially towards the film bulk [43] (see **Fig. 1)**:

$$\delta N(z) \approx \frac{N_0}{1 + \exp[-(z - h + h_0)/\Delta h]}. \tag{6}$$

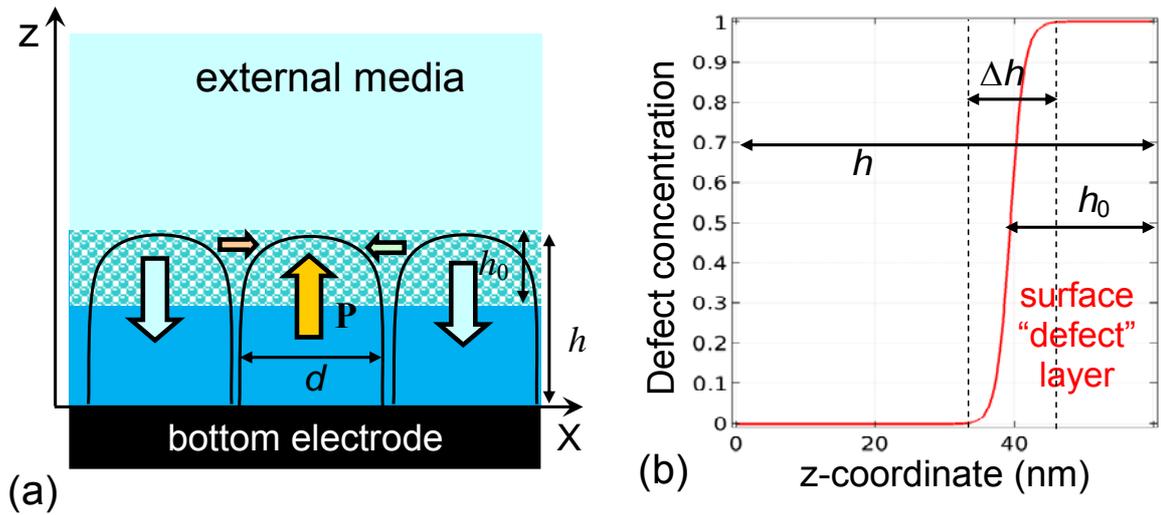

**FIGURE 1. (a)** Scheme of a film with the thickness $h$ and the layer of thickness $h_0$ where defects are accumulated. **(b)** Normalized concentration of defects inside the layer of thickness $h_0$ and transition layer depth $\Delta h$.

### III. RESULTS AND DISCUSSION

Using COMSOL Multiphysics package © we calculated ferroelectric polarization, electric fields and elastic properties from Eqs. (2-5) for exemplarily chosen film thickness, temperature, misfit strain, the defect distribution given by Eq. (6) and PbTiO$_3$ (**PTO**) ferroelectric parameters listed in **Table II**. Results of FEM calculations are shown in **Figs. 2-7.**

Table II. Description, dimension and numerical values of material parameters

| Description | Symbol and dimension | Numerical value for PbTiO$_3$ |
|---|---|---|
| Coefficient at $P^2$ | $\alpha(T)$ ($\times$C$^{-2}\cdot$m J) | $\alpha_T(T-T_C)$ |
| Inverse Curie-Weiss constant | $\alpha_T$ ($\times 10^5$C$^{-2}\cdot$m J/K) | 3.8 |
| Curie temperature | $T_C$ (K) | 752 |
| Background permittivity | $\varepsilon_b$ | 7 |
| Surface energy coefficient | $\alpha_{S0}$ ($\times$C$^{-2}\cdot$J) | 0 |
| Electrostriction coefficient | $Q_{ij}$ ($\times$m$^4$/C$^2$) | $Q_{11}$=+0.89, $Q_{12}$=−0.026, $Q_{44}$=0.0675 |
| Elastic stiffness tensor | $c_{ij}$ ($\times 10^{10}$ Pa) | $c_{11}$=17, $c_{12}$=8, $c_{44}$=11 |
| Elastic compliance tensor | $s_{ij}$ ($\times 10^{-12}$ 1/Pa) | $s_{11}$=8, $s_{12}$=−2.5, $s_{44}$=9 |
| Gradient coefficient | $g_{ij}$ ($\times 10^{-10}$C$^{-2}$m$^3$ J) | $g_{11}$=4.0, $g_{12}$=−0.5, $g_{44}$=0.5, |
| Flexoelectric stress tensor | $f_{ij}$ (V) | $f_{11}$=6.8, $f_{12}$=4.9, $f_{44}$=5.6 |
| Flexoelectric strain tensor | $F_{ij}$ ($\times 10^{-11}$m$^3$/C) | $F_{11}$=3, $F_{12}$=1, $F_{44}$=5 * |
| Kinetic coefficient | $\Gamma$ ($\times$s m/F) | 100 |
| LGD-coefficient at $P^4$ | $a_{11}$ ($\times 10^8$ JC$^{-4}\cdot$m$^5$) | −0.73 |
| LGD-coefficient at $P^6$ | $a_{111}$ ($\times 10^8$JC$^{-6}\cdot$m$^9$) | +2.60 |
| Surface screening length | $\lambda$ ($\times 10^{-10}$m) | 1 (or vary within the range) |
| Vegard strain coefficient | $W$ ($\times 10^{-30}$m) | ±10 |
| Misfit strain | $u_m$ (%) | −1 |
| Maximal defect concentration | $N_0$ ($\times 10^{26}$ m$^{-3}$) | (0 – 3) |

| Defect layer thickness | $h_0$ (nm) | 25 |
| Transition layer depth | $\Delta h$ (nm) | 1 |

\* The coefficients $F_{ij}$ are still not available experimentally for PTO, but some components could be evaluated from the first principles for various perovskites [44, 45, 46,] and their thin films [47]. On the other hand, the magnitudes of $F_{11}=3$, $F_{12}=1$ and $F_{44}=0.5$ (in $10^{-11}\text{C}^{-1}\text{m}^3$ units) are of the same order as the microscopic estimations ($F \sim 10^{-11}\text{m}^3/\text{C}$) by Kogan, and the values measured for SrTiO$_3$ by Zubko et al. [48, 49]. The value $F_{44}=5\times10^{-11}\text{C}^{-1}\text{m}^3$ is higher than a conventional one, but its effect is only relevant for fine details of polarization and elastic field distributions close to the bottom electrode. We note also that all values we used are significantly smaller than the ones ($F \sim (5-10)\times10^{-10}\text{m}^3/\text{C}$) measured for PbZrTiO$_3$ by Ma and Cross [50].

**A. Polarization, domain structure and elastic field dependence on the sign of Vegard coefficient**

Note, that in most cases, a stable poly-domain structure with prevailing out-of-plane polarization has been found for an applied negative misfit strain $u_m = -1\%$ and $\lambda > 0.1$nm [30]., which support the out-of-plane polarization component [7] and a poly-domain structures. The appearance of the closure domains [19] under the electrically open film surface depends strongly on the degree of screening, represented by the values of the surface screening length $\lambda$ and temperature [30].

To illustrate the above mentioned issues, **Fig. 2** shows the spatial distributions of the in-plane and out-of-plane polarization components, $P_x$ and $P_z$, respectively, corresponding elastic strains $u_{xx}$ and $u_{zz}$, and von Mises stress

$$\sigma_v = \sqrt{\left(\sigma_{xx}-\sigma_{yy}\right)^2+\left(\sigma_{yy}-\sigma_{zz}\right)^2+\left(\sigma_{zz}-\sigma_{xx}\right)^2+6\sigma_{yz}^2+6\sigma_{zx}^2+6\sigma_{xy}^2} \qquad (7)$$

in the cross-section of a 50-nm thick PTO film. The top and bottom rows are calculated for positive and negative Vegard coefficients, $W = +10\,\text{Å}$ and $W = -10\,\text{Å}$, respectively. At elevated temperature T=600 K, which, however, is sufficiently far from the film's transition temperature to the paraelectric phase, shallow (up to 5 nm) closure a-domains appear near the electrically open surface. They have a form of rounded wedges and relatively diffuse domain walls [see **Fig. 2(a)** and **2(f)** showing $P_x$ distribution]. There are clearly visible stripe c-domains with relatively sharp domain walls in the middle of the film and near the bottom screening electrode for the normal component of the polarization. The stripe domain walls noticeably broaden and diffuse to the depth of about 5 nm near the top surface [see **Fig. 2(b)** and **2(g)** showing $P_z$ distribution]. The polarization in the middle of the closure and stripe domains is significantly larger for the Vegard coefficient $W = +10\,\text{Å}$, than it is for $W = -10\,\text{Å}$, but all other characteristics of a- and c- domains depend weakly on the value of $W$ [compare **Fig. 2(a)** and **2(f), Fig. 2(b)** and **2(g)**]. A 25 nm layer of elastic defects, the domain structure, and the misfit strain at the film-substrate interface determine the structure and spatial distribution of the elastic strain tensor in the film, whose diagonal components $u_{xx}$ and $u_{zz}$ are shown in **Fig. 2(c), 2(d)** and **2(h), 2(i)**, respectively. The main features on the lateral strain distribution are caused by the domain structure via the piezoelectric and flexoelectric effects, and so the distribution of $u_{xx}$ is virtually independent on the sign of $W$ [compare **Fig. 2(c)** and **2(h)**]. The main features of the vertical strain distribution are conditioned not only by the domain structure, but also by an elastic field gradient in the defect layer. That is why a diffuse horizontal boundary is clearly visible on the edge of the defect layer in **Fig. 2(d)** and **2(i)**. The vertical strain in this layer is determined by a chemical pressure of defects and thus it changes sign when the sign of $W$ is changed [compare **Fig. 2(d)** and **2(i)**]. The distribution of von Mises dilatational stress $\sigma$ reproduces the profile of the out-of-plane polarization component, namely, a stripe domain structure with broadened domain walls near the surface while the value of $\sigma$, in a near-surface layer with a thickness of the order of 5 nm, is strongly dependant on the sign of $W$ [see **Fig. 2(e)** and **2(j)**]. We note that the pronounced features of the distributions of $u_{xx}$, $u_{zz}$ and $\sigma$ near the bottom electrode do not depend on the sign of $W$, since they arise from the flexoelectric coupling.

Note that the value of the screening length $\lambda$ strongly affects the polar properties of the film, determines its critical thickness at fixed temperature and the existence as well as the type of the domain structure [30]. In addition, a pronounced minimum at a certain width, which depends on $W$, temperature and film thickness, appears on the dependence of the system specific energy $E$ on the domain lateral size $d$ when $\lambda$ increases [see **Fig.3**]. The decreasing dependence $E(d)$ is steeper, and the minimum on it is much deeper for positive $W = +10\,\text{Å}$ than for negative $W = -10\,\text{Å}$ [compare **Figs. 3(a)** and **3(b)**].

Notably, the expected Kittel-Mitsui-Furuichi (**KMF**) relation connecting the period $d$ of the stripe domain structure having infinitely thin walls with the film thickness $h$, $d \sim \sqrt{h}$, is not confirmed in our calculations, since they naturally account for domain wall broadening near electrically-open surfaces (via the polarization gradient) and closure domains (via polarization rotation) [30]. Moreover, our results are $\lambda$- and $W$-dependent. To illustrate this, **Fig. 3(c)** shows the dependences of the equilibrium domain size $d$ on the screening length $\lambda$. The dependence on the domain size on $\lambda$ obeys an analytical formula, $d = d_0 + \dfrac{D}{\lambda - \lambda_{cr}}$ [see **Fig. 3(d)** showing the dependence of inverse value $1/(d - d_0)$ on $\lambda$], where the critical values $\lambda_{cr}$ slightly differ for $W = +10\,\text{Å}$ and $W = -10\,\text{Å}$, while the parameters $d_0$ and $D$ depend on the $W$ sign much more strongly [see caption to **Fig. 3**].

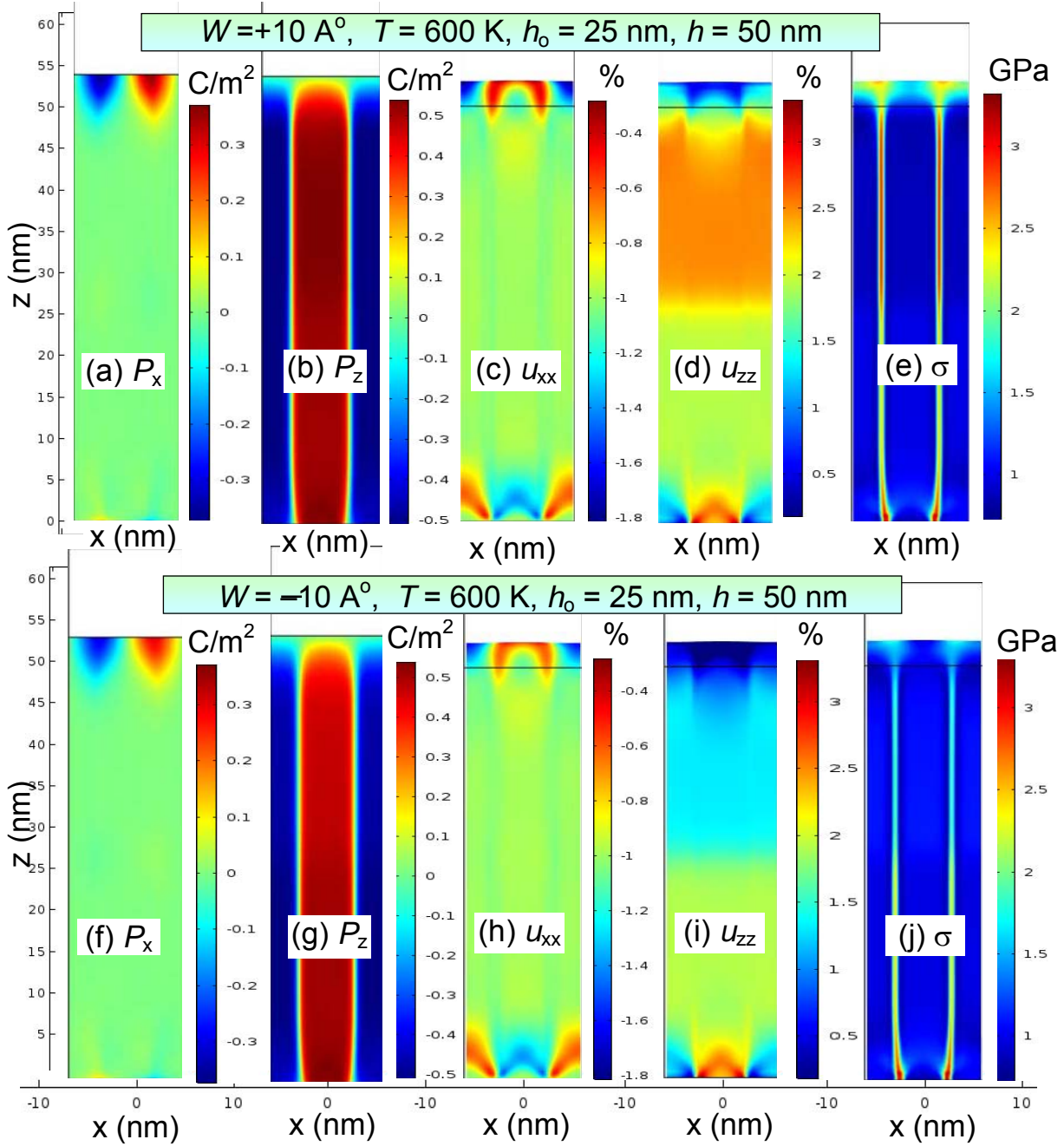

**FIGURE 2.** Spatial distribution of the in-plane polarization component $P_x$ **(a,f)** and the out-of-plane component $P_z$ **(b,g)**, elastic strains $u_{xx}$ **(c,h)** and $u_{zz}$ **(d,i),** and von Mises stress $\sigma$ **(e,j)** in the cross-section of the 50-nm PTO film calculated for positive [top row, plots **(a-e)**] and negative [bottom row, plots **(f-j)**] Vegard coefficients $W = \pm 10 \,\mathring{A}$, temperature $T$=600 K, screening length $\lambda$=0.1nm, depth of defect layer $h_0$=25 nm, $\Delta h$=1 nm, and defect concentration $N_0 = 3\times 10^{26}$ m$^{-3}$. Other parameters are listed in **Table II.**

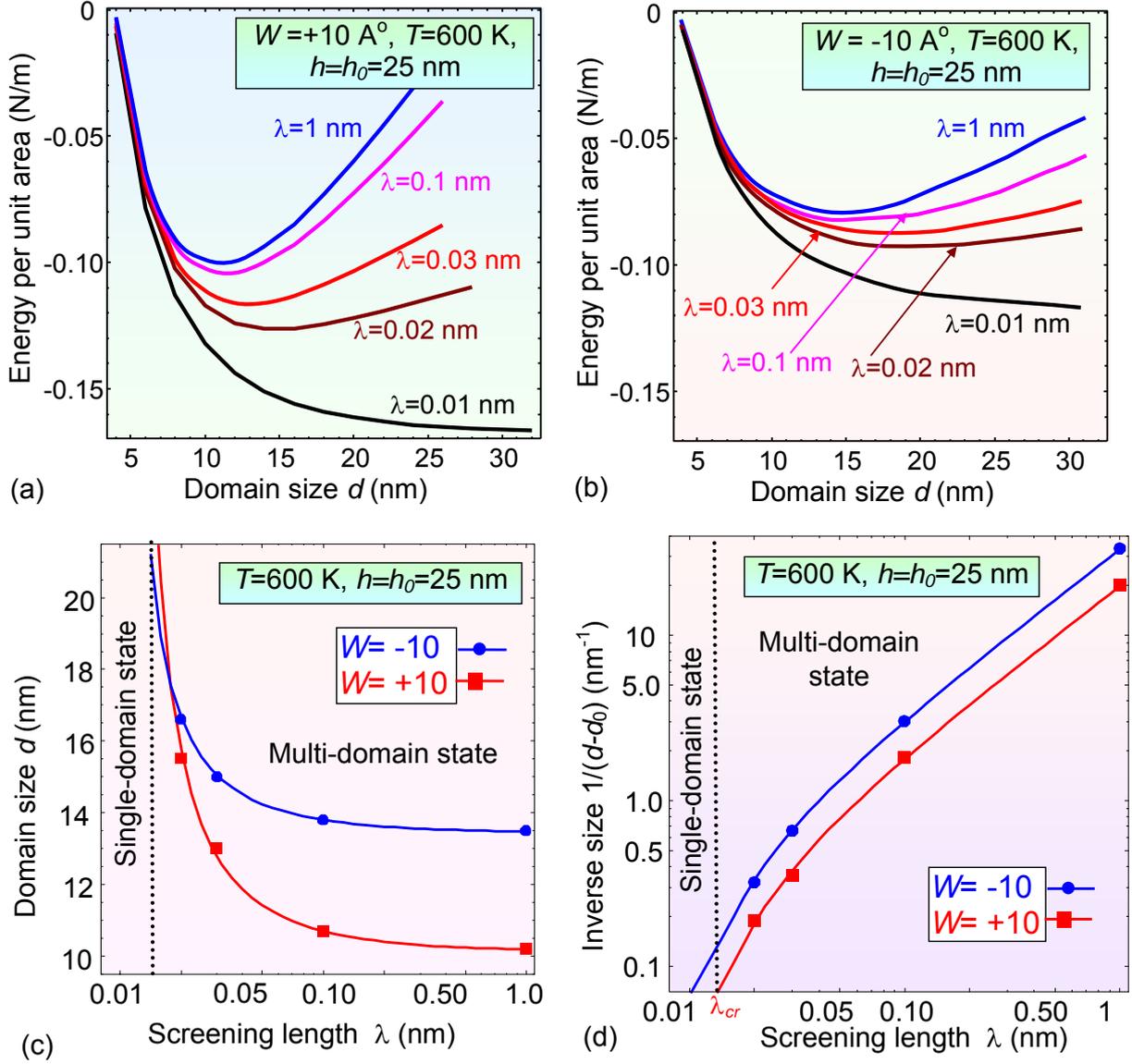

**FIGURE 3.** **(a)** The dependence of the 25-nm thick PbTiO$_3$ film total energy on the domain size (potential relief of wall-wall interaction) calculated for different values of surface screening length $\lambda$ (shown near the curves), positive Vegard coefficient $W = +10$ Å **(a)** and negative Vegard coefficient $W = -10$ Å **(b)**. Dependences of the equilibrium domain size $d$ **(c)** and inverse value $1/(d - d_0)$ **(d)** on the surface screening length $\lambda$ calculated numerically for $W = -10$ Å (circles) and $W = +10$ Å (squares). Solid curves are fitting to the formula $d = d_0 + \dfrac{D}{\lambda - \lambda_{cr}}$, where $d_0 = 13.5$ nm, $D = 0.03$ nm$^2$, $\lambda_{cr} = 0.010$ nm for $W = -10$ Å; and $d_0 = 10.2$ nm, $D = 0.05$ nm$^2$ and $\lambda_{cr} = 0.011$ nm for $W = +10$ Å. Temperature T=600 K and defect concentration $N_0 = 3 \times 10^{26}$ m$^{-3}$. Other parameters are listed in **Table II.**

**B. Temperature evolution of spontaneous polarization, domain structure and elastic fields**

**Figure 4** shows the temperature dependencies of the maximum spontaneous polarization at the center of the stripe domains, calculated for films of different thicknesses (6 – 50) nm with a layer of elastic defects (solid curves) and without it (dashed curves). Note that the average polarization is zero due to the presence of stripe domain structure. The temperature of spontaneous polarization and domain structure appearance in a film with defects is much larger (by 50 – 70 K), than for films without them, and the polarization itself is somewhat larger for thin films with a thickness less than 25 nm, for which the defect layer occupies the whole film and the Vegard effect is positive ($W = +10 \text{ Å}$) [compare solid and dashed curves in **Fig. 4(a)**]. When increasing the film thickness to 50 nm (with a thickness of the defect layer 25 nm), the temperature of the spontaneous polarization emergence becomes 20 K higher than the ferroelectric transition temperature of a 50-nm film without defects [compare solid and dashed curves in **Fig. 4(b)**]. Notably, the bending appears on the temperature dependence of the maximum polarization at the temperature 550 K, being related with the emergence of closure domains at lower temperatures. The temperature of polarization emergence decreases at negative $W = -10 \text{ Å}$ (this case is not shown in the figures, since we are primarily interested in the conditions of polar properties enhancement).

Spatial distributions of in-plane and out-of-plane polarization components and corresponding elastic strains in the cross-section of the 60-nm thick PTO film calculated for positive Vegard coefficients $W = +10 \text{ Å}$, elevated (850 K) and room (300 K) temperatures are shown in **Fig.5**. It is evident that the closure domains, as well as a pronounced stripe domain structure, are absent at high temperatures near the phase transition of the film into the paraelectric phase [compare **Figs. 5(a)-(d)** and **5(e)-(h)**]. On the contrary, small domains, which branch near the surface of the film, appear at 850 K. They gradually "freeze" and transform to stripe domain structure with closure domains as the temperature decreases.

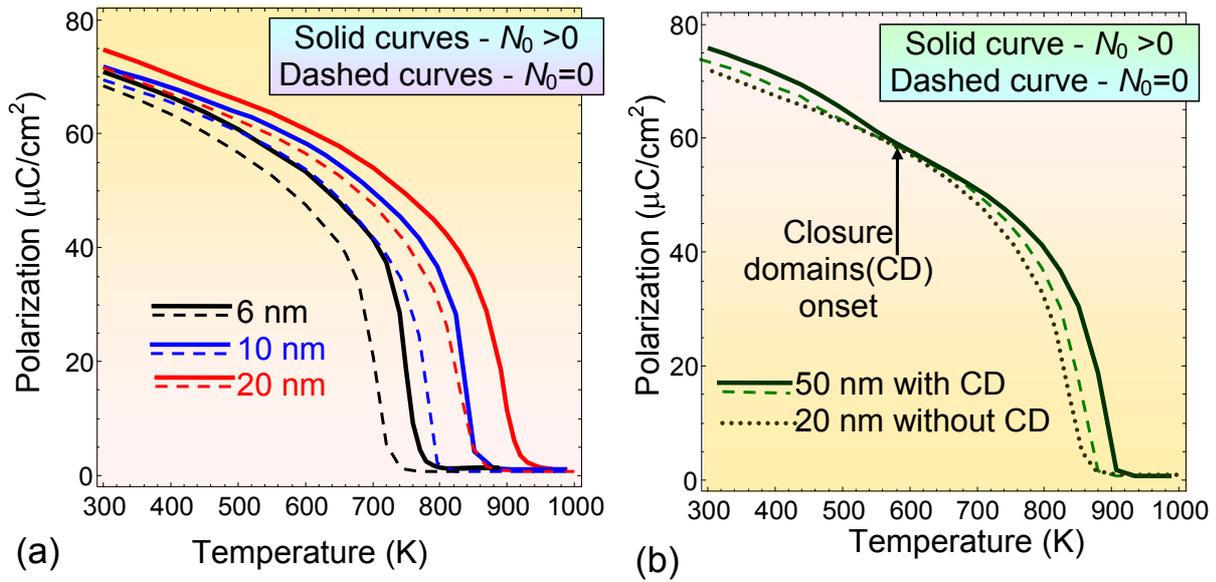

**FIGURE 4.** Temperature dependence of the maximal spontaneous polarization calculated for different film thicknesses $h$ = (6, 10, 20) nm [plot **(a)**] and $h$ = (20, 50) nm [plot **(b)**], without defects ($N_0 = 0$, dashed curves) and with defect concentration $N_0 = 2\times10^{26}$ m$^{-3}$ and Vegard coefficient $W = +10$ Å [plot **(b)**], screening length λ=0.1nm. Other parameters are listed in **Table II**. The inflections at the curves for 50-nm thick film indicate the appearance of the closure domains (CD) at temperatures lower than 550 K.

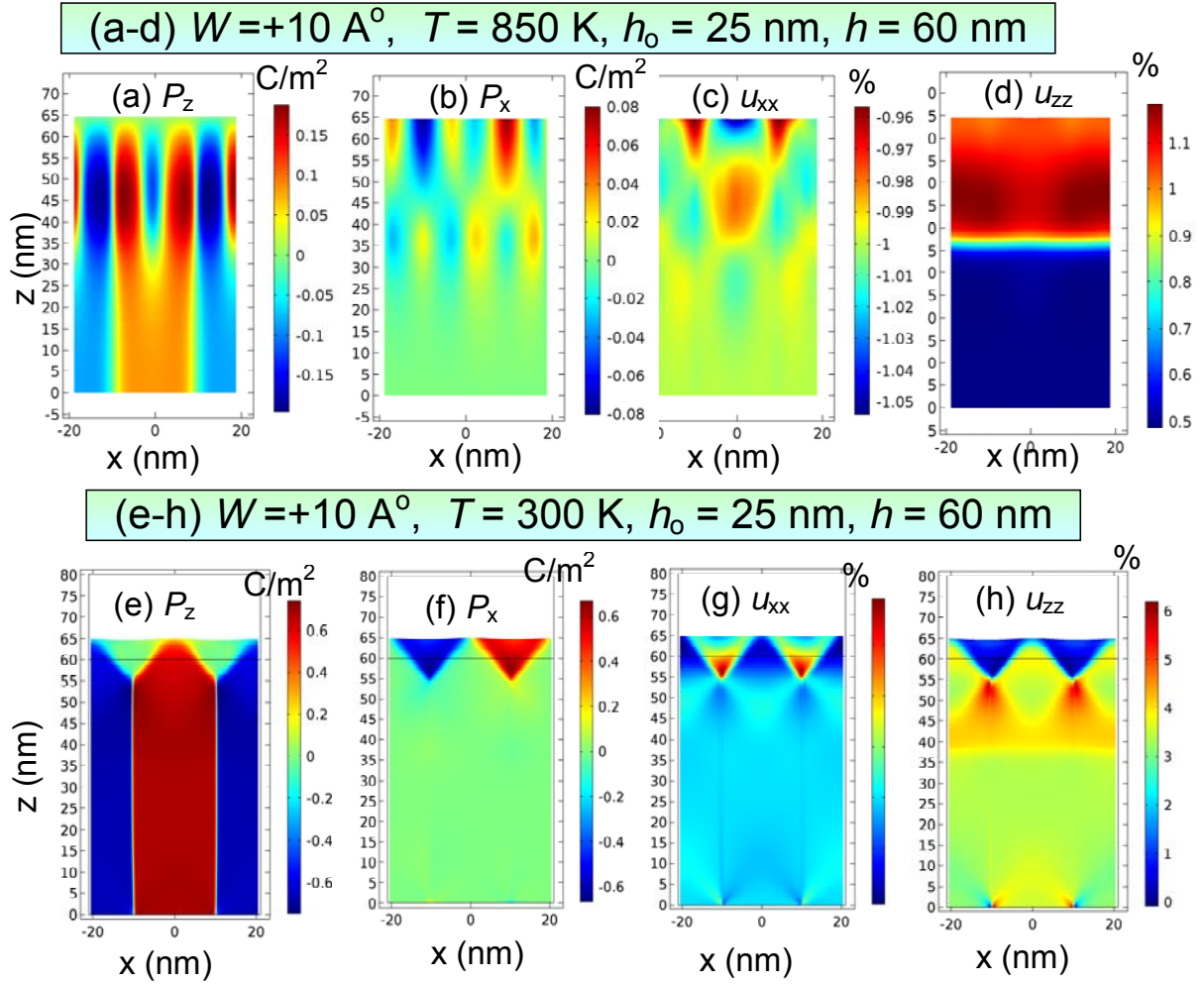

**FIGURE 5.** Spatial distributions of polarization components $P_x$ **(a, e)** and $P_z$ **(b, f)**, elastic strains $u_{xx}$ **(c, g)** and $u_{zz}$ **(d, h)** in the cross-section of the 60-nm thick PTO film calculated for positive Vegard coefficients $W = +10\,\overset{o}{\text{A}}$, temperatures T=850 K [plots **(a)-(d)**] and 300 K [plots **(e)-(h)**], screening length λ=0.1nm and defect concentration $N_0 = 3\times 10^{26}$ m$^{-3}$. Other parameters are listed in **Table II**.

### C. Ferroelectric transition temperature dependence on the film thickness

**Figure 6** shows the dependence of the ferroelectric transition temperature $T_C(h)$ on the film thickness $h$ [**Fig. 6(a)**] and its inverse value $1/h$ [**Fig. 6(b)**], calculated for the positive, zero and negative Vegard coefficients $W$. At $W > 0$ the maximum appears on the dependence at a film thickness of 25 nm, virtually equal to the thickness of the defect layer $h_0$ [see red curve in **Fig. 6(a)**]. Temperature $T_C(h)$ decreases monotonically with decreasing $h$ at $W \leq 0$. Notably, the inequality $T_C(h, W < 0) < T_C(h, W = 0) < T_C(h, W > 0)$ is valid for thicknesses more than 4 nm [compare red, magenta and blue curves in **Fig. 6(a)**]. At a film thickness of about 3.5 nm, all three curves intersect, and the order of the curves corresponding to $W > 0$ and $W < 0$ changes with the further decrease of the film thickness. A maximum and a kink on dependencies $T_C(h)$ are observed at $h = h_0$, for positive and negative $W$, respectively [see red and blue curves in **Fig.6(a)**].

From **Figure 6(a)** we can see, that the critical thickness of the film below which the ferroelectric phase vanishes is absent for all $W$. It is true even until 2 nm thickness (that is about 5 lattice constants) for which the continual theory of LGD is still applicable at least qualitatively. Somewhat overstretching the LGD approach one can observe in **Figure 6(b)** that $T_C$ does not decrease below room temperature for the films with thickness $h \geq 1$ nm or even less, its value varying within the range (350-450) K in dependence on the sign and value of $W$ [compare red, magenta and blue curves with symbols in **Figs. 6**]. This effect can be explained only by the presence of relatively strong compressive strains (-1%) at the film-substrate interface, which effectively support spontaneous dipole displacements in ulta-thin films [3-5] due to the electrostriction [7] and flexoelectric effect [28]. The depolarization field in the film is minimal due to a developed domain structure [see **Figs. 2** and **4**]. Indeed, electrostrictive coupling between polarization and elastic stresses shifts the transition temperature significantly, in the order of $Q_{33ij}\sigma_{ij}/\alpha_T$ in a stressed film (see ref. [7] for details), and the flexoelectric effect creates a built-in electric field proportional to the convolution of tensors $F_{jkim}\sigma_{jk}n_m$ in the boundary conditions Eq. (2b) (see ref.[28] for details). The approximate part of the dependence $T_C(h)$ calculated analytically for small thicknesses without the flexoelectric effect and Vegard effect is shown in **Fig. 6 (b)** by a dotted line.

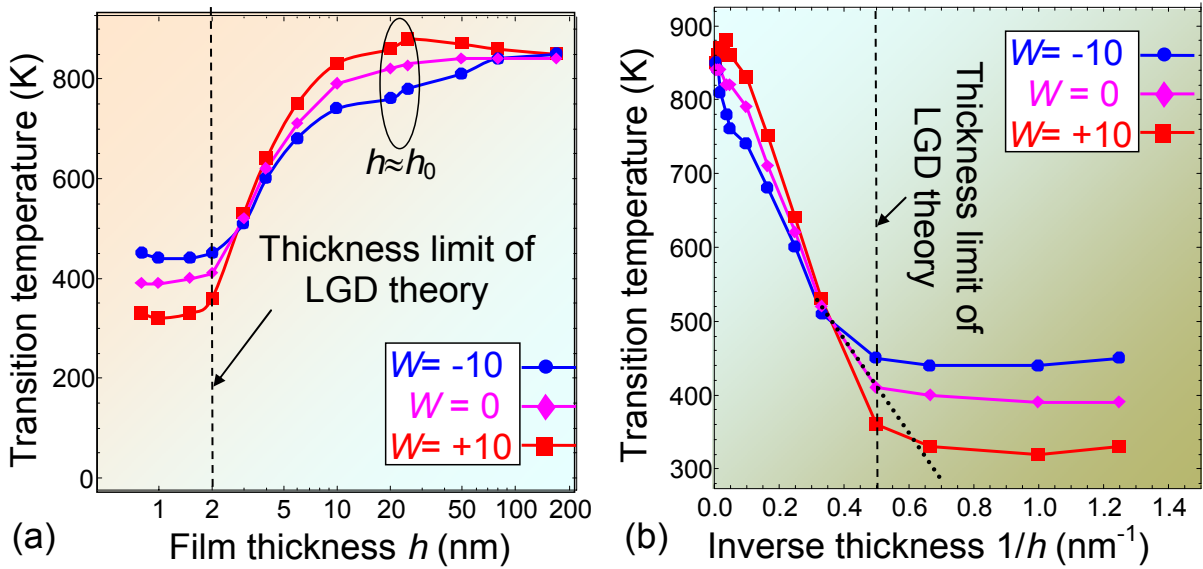

**FIGURE 6.** Ferroelectric transition temperature vs. the film thickness $h$ **(a)** and $1/h$ **(b)**. Screening length $\lambda$=0.1nm, $N_0 = 3\times10^{26}$ m$^{-3}$, Vegard coefficient $W = +10$ Å (squares), $W = 0$ (diamonds) and $W = -10$ Å (circles), and nonzero flexocoupling. Dotted line in plot (b) corresponds to the case $W = F = 0$. Vertical dashed lines indicate the thickness limit of continuum LGD-theory applicability. Other parameters are listed in **Table II.**

Distributions for the 2-nm film calculated at the temperature 300 K show the film in the state close to the phase transition; that is why its polarization is severely weakened and domain walls are notably diffused. Also, a metastable domain state can be observed for the film at negative $W$. This illustrates sensitivity of thin films to lateral boundary conditions and flexoeffect that carry a major responsibility for the formation of such kinds of structures. Flexoelectric coupling, in particular, is also responsible for the very existence of the ferroelectric phase in thin films under 6 nm.

A possible explanation for an anomalous change in the phase-transition curve evolution at $h<4$ nm [that is shown in **Fig. 6(a)**] could be a transition from the c-domain state of the film with polarization perpendicular to the surface in thick films (where a part of closure domains is relatively small because of their localization at approximately 5 nm below the surface, see **Figs. 2** and **5**) to the mainly a-domain state with the decrease of thickness, owing to the flexocoupling. Indeed, with the thickness decrease a-domains with the in-plane polarization direction become significant (see **Fig. 7**). This happens because it is well known that compressive misfit strains $u_m<0$ support the c-domain formation, while the dilatation ones $u_m>0$ support the a-domain formation. Respectively, $W>0$ supports c-domain stability, while $W<0$ supports the stability of a-domains. Therefore, for the case of thickness decrease below 4 nm in the film already having defects homogeneously occupying its whole bulk (because $h_0=25$ nm $>>$ 4 nm), it is energetically favorable to increase the fraction of a-domains, so that the ferroelectric phase transition temperature for this scenario is higher. This can be seen in **Fig. 6**; the detailed analysis of the corresponding domain structure and elastic fields for the film thicknesses below 5 nm is given in **Fig. 7**.

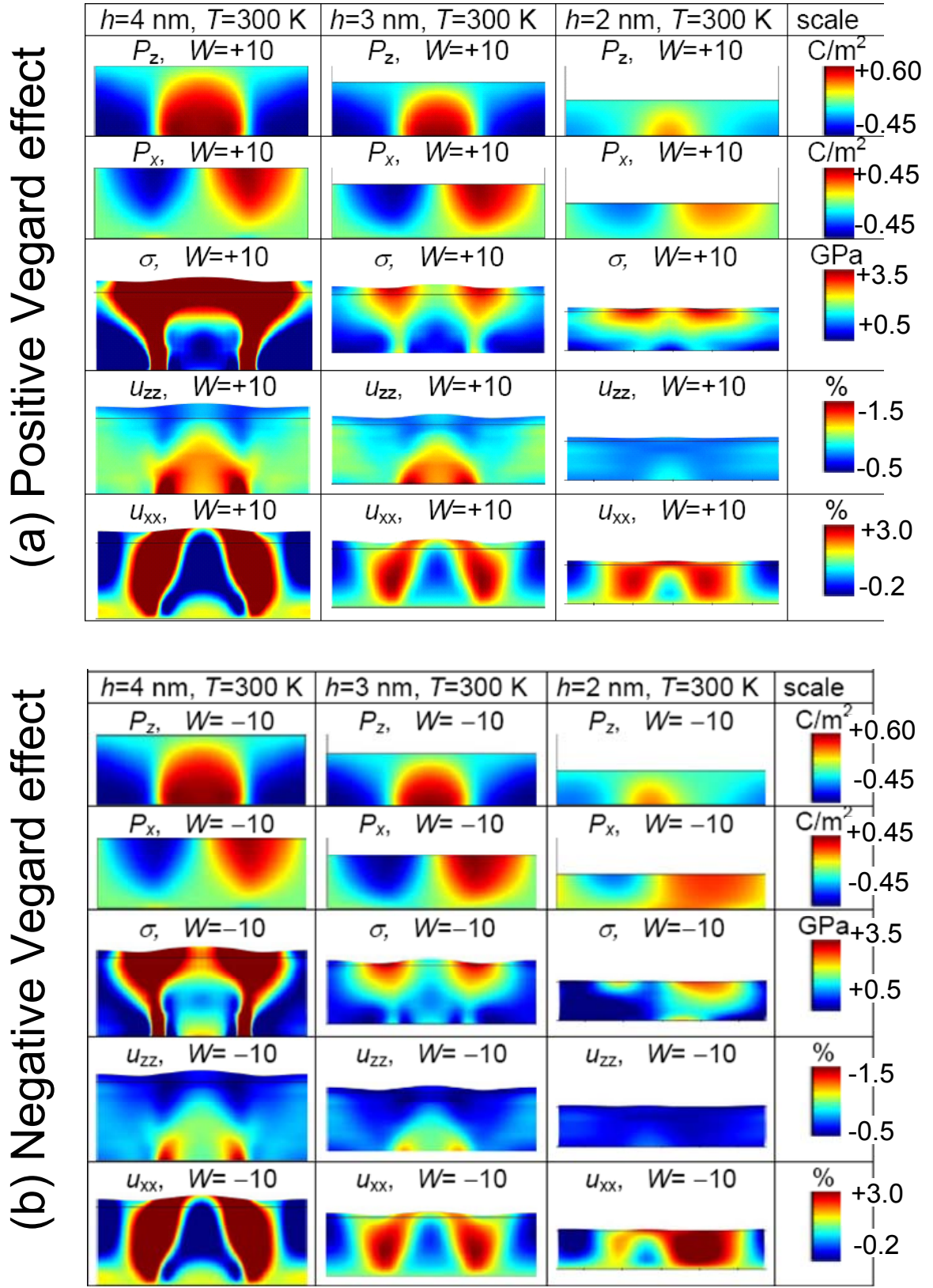

**FIGURE 7.** Spatial distribution of the out-of-plane and in-plane polarization components $P_x$ and $P_z$, von Mises stress $\sigma$ and elastic strains $u_{xx}$ and $u_{zz}$ in the cross-sections of the 4, 3 and 2-nm PTO film calculated for positive [top part **(a)**, $W = +10$ Å] and negative [bottom part **(b)**, $W = -10$ Å] Vegard coefficients, room temperature

T=300 K, screening length λ=0.1nm, and defects filling the entire film with a concentration $N_0 = 3\times 10^{26}$ m$^{-3}$. Other parameters are listed in **Table II.** Color gradient denotes scales of for the following physical parameters. In-plane polarization changes from −0.45 to 0.45 C/m²; out-of-plane polarization changes from −0.45 to 0.6 C/m²; Von Mises stress changes from 0.8 to 3.6 GPa; in-plane strain changes from −1.7% to −0.3%; out-of-plane strain changes from −0.2% to 3.1%.

Polarization components change their behavior when the film thickness approaches 4 nm and less. The out-of-plane component that prevailed in thicker films gradually dims, giving the place for the in-plane polarization rising amplitude and growing area of closure domains. 3-nm film is already seen as such where the a-domains slightly exceed the c-domains in size. While the sign of the Vegard effect coefficient has an insignificant effect on the polarization amplitude and domain shapes, it can change mechanical strain and stress distributions. It can be seen in **Fig. 7** that there are different elastic fields, specifically the out-of-plane component $u_{zz}$, changing significantly under positive and negative *W*, which can be consistently traced back to the thicker films (see **Fig. 2**), where such dependencies occur in the defect-rich part of a ferroelectric bulk. Since defects are quasi-uniformly spread across the depth of the thin film ($h_0 >> h$), Vegard stresses impact the whole film thickness. Note that the asymmetry of the out-of-plane polarization scale (from −0.45 to 0.6 C/m²) at film thickness 2-3 nm originates from the built-in electric field induced by flexo-chemical coupling, and the asymmetry is absent for thicker films [compare **Figs.7** with **Figs. 2** and **5**]. Temperature dependence of the total energy per unit area in the films of thickness (2 − 170) nm is shown in Appendix A.

Graphs in **Figs. 6-7** are plotted for the fixed concentration of defects. Their detailed analysis, carried out for various defect concentrations in the temperature range of (600 − 900) K with a positive Vegard coefficient, shows, that a pronounced maximum appears on the transition temperature dependence $T_C(h)$ at $h \approx h_0$ with the increase of the defect concentration [see **Fig. 8(a)** in a semi-logarithmic graph]. At the maximum, the transition temperature of (20-30) nm film with a layer of defects near the surface exceeds by 50 K the transition temperature of a thick lead titanate film, which makes it possible to significantly improve the polar properties of thin films. The transition temperature of (20-30) nm film without defects is about 200 K lower than the one in the film with defect concentration $3\times 10^{26}$ m$^{-3}$ and $W = +10 \overset{\text{o}}{\text{A}}$. Note that we neglected the relaxation of the mismatch deformations in films with thickness $h > h_0$, and therefore the applied compressive strain (-1%) leads to renormalization of the Curie bulk temperature from 752 K to 880 K in 100-nm films.

Dependence of transition temperature on the defect concentration $N_0$ increases quasi-linearly at $W > 0$, and its slope increases with the film thickness decrease [**Fig. 8(b)**]. Solid curves with empty symbols in **Fig. 8(a)** and dashed curves in **Fig. 8(b)** calculated accounting for the flexoelectric effect with flexoelectric coefficient $F_{ij} > 0$ listed in **Table I** correspond to a higher $T_C(h)$ than the curves calculated at $F_{ij} = 0$. The difference is most significant for the thinnest films [see the curves for $h$=10 nm and 20 nm in **Fig. 8(b)**]; it decreases with the film thickness increase and is almost nondescript for the films with a thickness in the order of 100 nm and above [see the curves for $h$=80 nm and 170 nm in **Fig. 8(b)**]. This is obviously related with the built-in electric field induced by the flexoelectric coupling of polarization with inhomogeneous elastic stresses, which is proportional to the product $F_{jkim}\sigma_{jk}n_m$ [see boundary conditions (2b)]. Note that the transition temperature substantially rises with defect concentration increase at positive Vegard coefficient (with the flexoeffect or without it), but the reentrant ferroelectric phase, observed experimentally in spherical nanoparticles with a radius $R < 5$ nm [18] and then explained theoretically by flexo-chemical coupling [17], has not been observed in thin films. In our opinion this is because there is no curved surface in the films that induces the ferroelectric phase due to a competition between the contributions of size effects and surface tension into the Curie temperature shift, which have different signs and are proportional to $1/R$ and $1/R^2$, respectively [17].

The pronounced maximum on the transition temperature contour maps in the variables "film thickness - defect concentration" exists under the presence of flexoelectric coupling [see **Fig. 8(c)**] and without it [see **Fig. 8(d)**], however, the flexoelectric effect significantly shifts the transition temperature (by up to 30 K for thin PbTiO$_3$ films).

Thus, the position and height of the maximum $T_C(h)$ can be controlled by the defect concentration in the layer and the surface screening length, which can be useful for advanced applications. Summarizing this section, we conclude that uncharged elastic defects have an unexpectedly strong impact on the polar and elastic properties of ferroelectric films due to Vegard stresses in the defective layer of the film and the flexoelectric effect.

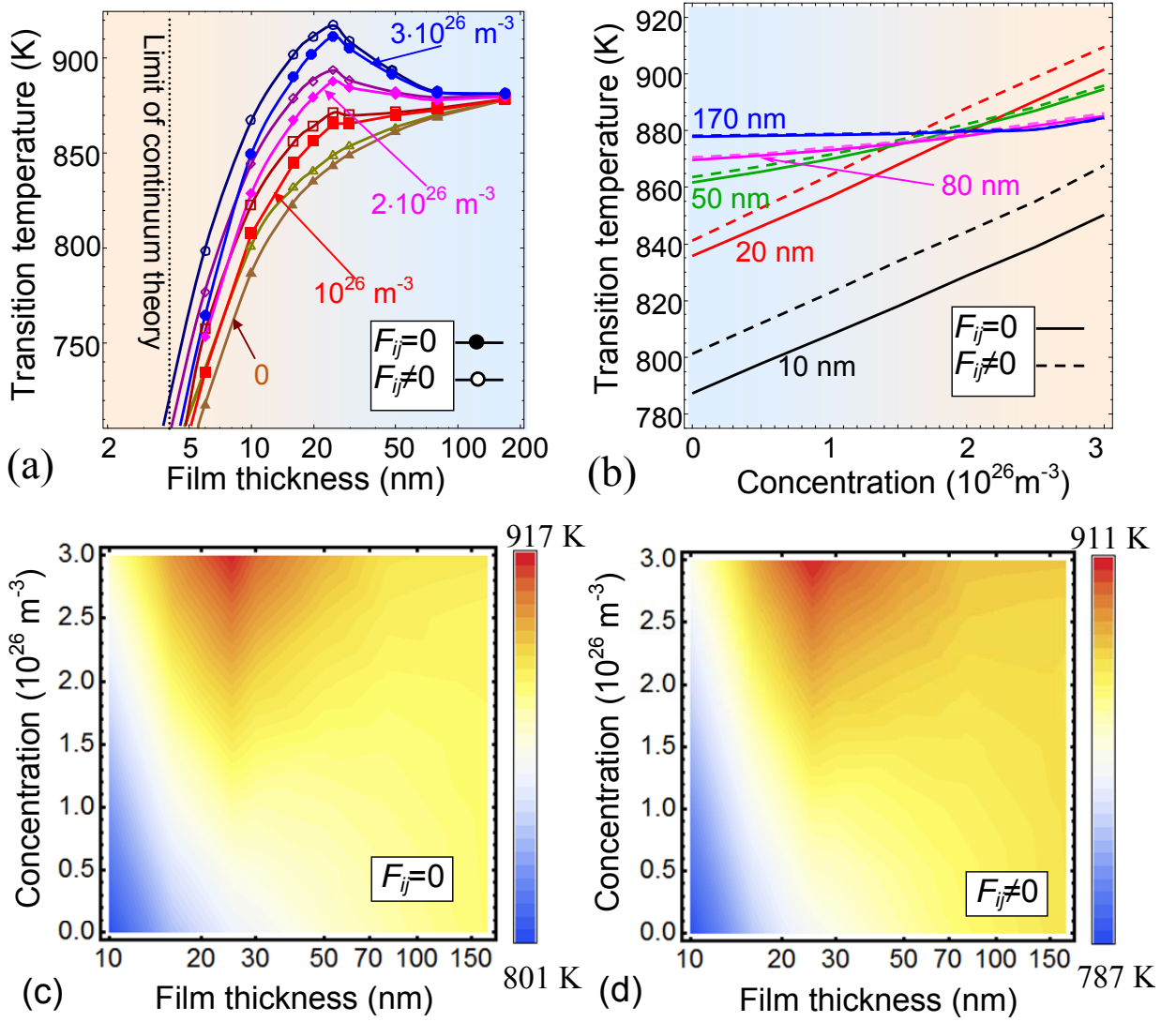

**FIGURE 8.** Ferroelectric transition temperature dependence on the film thickness and defect concentration. **(a)** Dependence of the transition temperature on the film thickness calculated for the different values of the defect concentration $N_0$ (shown near the curves), zero (empty symbols) and nonzero (filled symbols) flexoelectric coefficients. **(b)** Dependence of the transition temperature on defect concentration calculated for different values of the film thickness (shown near the curves); zero (solid curves) and nonzero (dashed curves) flexoelectric coefficients. **(c, d)** Contour maps in the coordinates "film thickness – defect concentration" for the two cases of zero **(c)** and nonzero **(d)** flexoelectric coefficients. Room temperature, screening length λ=0.1nm, $W = +10$ Å, and the depth of defect layer $h_0$=25 nm, Δh=1 nm,. Other parameters are listed in **Table II.**

## IV. CONCLUSIONS

Using Landau-Ginzburg-Devonshire approach we established the effect of the flexoelectro-chemical coupling on the polar properties and phase transitions in thin ferroelectric films with a surface layer of uncharged elastic point defects (vacancies or ions). We considered a typical case, when the defects are concentrated in a thin layer below the top film surface creating a sharp gradient of elastic fields. The defective surface of the film is not covered with an electrode, but with an ultra-thin layer of ambient screening charges, which are characterized by a surface screening length.

We obtained that the influence of the flexoelectro-chemical coupling and surface screening length on the ferroelectric transition temperature of the film, distribution of the spontaneous polarization and elastic fields, domain wall structure and period is rather strong, namely, it turned out that:

- The screening length strongly affects the polar properties and domain structure in the film. In particular, a pronounced minimum appears on the dependence of the system's specific energy on the domain size with an increase of the screening length, the depth of the minimum depending essentially on the magnitude of the Vegard coefficient.

- Due to the flexoelectric effect there is no size-induced transition to a paraelectric phase until (2 – 4) nm thickness of $PbTiO_3$ films with 1% of compressive misfit strain. The origin of this phenomenon is the re-building of the domain structure in the film (namely the cross-over from c-domain stripes to a-type closure domains) emerging with its thickness decrease below 4 nm, conditioned by the flexoelectric coupling and facilitated by negative Vegard coefficient. Though we observe no phase transition for smaller thickness, our results (as obtained in the continuum theory framework) can be inaccurate below the (2-4) nm size. Despite the said limitation the obtained results point at tempting opportunities for defect-strain engineering of the ultra-thin perovskite film ferroelectric properties and domain structure tuning, which can be very promising for the ferroic film applications in nanoelectronics.

- Electric field induced by the defect layer has an unexpectedly strong influence on the polar and elastic properties of the strained films due to the coupling of inhomogeneous Vegard stresses and the flexoelectric effect (defect-driven flexo-chemical effect). Positive Vegard coefficients and high concentration of elastic defects effectively maintain the ferroelectric transition temperature above 350 K in the strained $PbTiO_3$ films due to the flexo-chemical effect. In contrast to the pure flexoelectric effect coefficients, which values are material-specific constants, the magnitude of the flexo-chemical effect can be controlled by the concentration of defects, their type and distribution in the film, making the considered system much more suitable for tuning.

- The increase of defect concentration leads to a noticeable monotonic decrease in the ferroelectric transition temperature of the $PbTiO_3$ film with negative Vegard coefficients. In contrast, for positive Vegard coefficients, a pronounced maximum (with a height up to 200 K) appears on the thickness dependence of the transition temperature with increasing defect concentration. The film thickness corresponding to the maximum is approximately equal to the thickness of the defect layer and relatively weakly depends on the surface screening length. The latter property may have important implications for miniaturization of ferroelectric devices.

- The pronounced maximum on the dependence of the ferroelectric transition temperature on the film thickness exists even without the flexoelectric coupling in the film, however, the coupling strongly shifts the transition temperature (by up to 30 K for thin $PbTiO_3$ films). Since the maximum position and height can be controlled by modifying the defect concentration and Vegard coefficient, the obtained results are promising for advanced applications in ferroelectric memory devices and those applications in nanoelecronics, where introducing of different types and amounts of defects is conceivable.

**Acknowledgements**

S.V.K study was supported by the U.S. DOE, Office of Basic Energy Sciences (BES), Materials Sciences and Engineering Division (MSED) under FWP Grant No. A portion of this research was conducted at the Center for Nanophase Materials Sciences, which is a DOE Office of Science User Facility. I.S.V. gratefully acknowledges support from the Deutsche Forschungsgemeinschaft (DFG) through the grant GE 1171/7-1.

**Authors contribution**

E.A.E. wrote the codes and, assisted by I.S.V. and E.N.F., performed numerical calculations. A.N.M. stated the problem and wrote the manuscript draft. E.A.E. and I.S.V. prepared the figures. Results analyses and manuscript improvement was performed by Y.A.G. and M.D.G. and S.V.K.

# APPENDIX

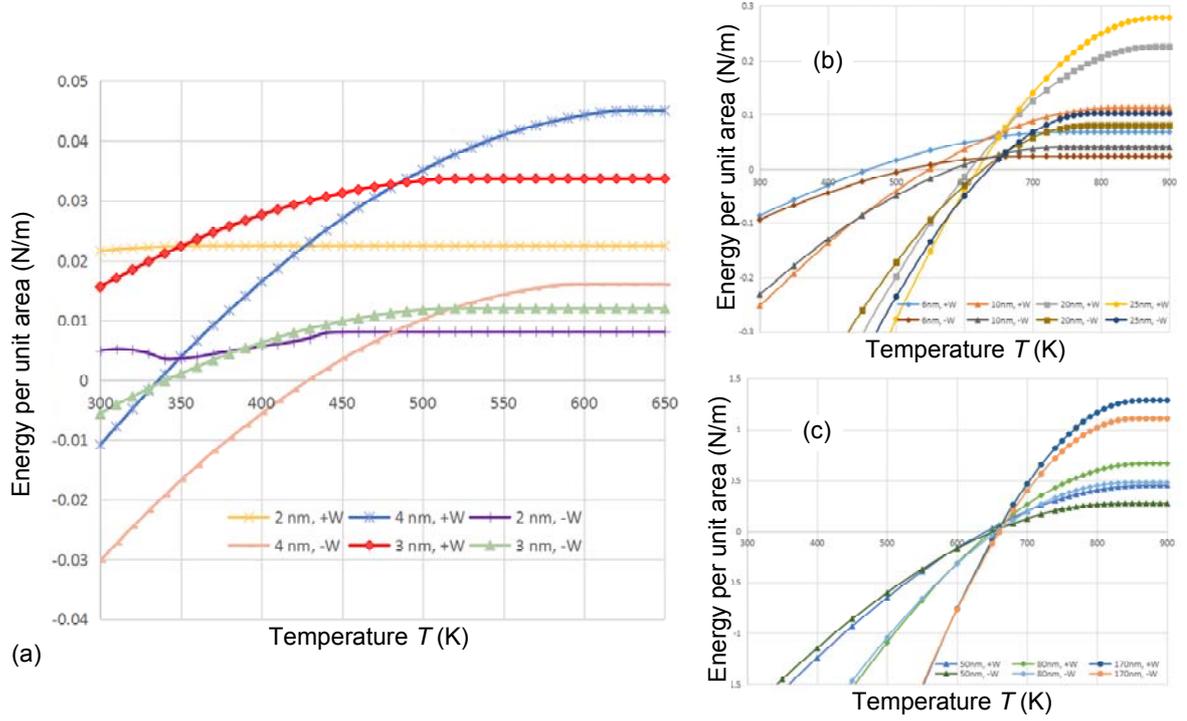

**FIG. A1.** Temperature-dependent energy diagram for the total energy in the thin films of thickness 2, 3 and 4 nm (a), 6, 10, 20 and 25 nm (b), 50, 80, 170 nm (c) with the defect concentration $N_0=3\times10^{26}$ 1/m³, $h_0=25$ nm, $\Delta h=1$ nm, $\lambda=0.1$ nm, at the room temperature (300 K), and the positive ($W=+10$ Å) and negative ($W=-10$ Å) Vegard coefficients. Saturation plateaus on the curves correspond to the paraelectric phase. Linear or parabolic-like dependencies correspond to the ferroelectric phase in the vicinity of respectively I and II-type phase transition. Note that the total energy of the system is greater than zero because of its mechanical component modified by a misfit strain.